\documentclass[twocolumn,aps,superscriptaddress]{revtex4}

\usepackage{amssymb}
\usepackage{amsmath}
\usepackage{graphicx}
\usepackage[normalem]{ulem}
\usepackage[dvips]{color}
\usepackage{appendix}

\begin{document}

\title{Constraining isovector nuclear interactions with giant resonances within a Bayesian approach}

\author{Jun Xu\footnote{xujun@zjlab.org.cn}}
\affiliation{Shanghai Advanced Research Institute, Chinese Academy of Sciences,
Shanghai 201210, China}
\affiliation{Shanghai Institute of Applied Physics, Chinese Academy
of Sciences, Shanghai 201800, China}
\author{Jia Zhou}
\affiliation{Shanghai Institute of Applied Physics, Chinese Academy
of Sciences, Shanghai 201800, China}
\affiliation{University of Chinese Academy of Sciences, Beijing 100049, China}
\author{Zhen Zhang}
\affiliation{Sino-French Institute of Nuclear Engineering and Technology, Sun Yat-Sen University, Zhuhai 519082, China}
\author{Wen-Jie Xie}
\affiliation{Department of Physics, Yuncheng University, Yuncheng 044000, China}
\author{Bao-An Li}
\affiliation{Department of Physics and Astronomy, Texas A$\&$M University-Commerce, Commerce, TX 75429, USA}

\date{\today}

\begin{abstract}
We put a stringent constraint on the isovector nuclear interactions in the Skyrme-Hartree-Fock model from the centroid energy $E_{-1}$ of the isovector giant dipole resonance in $^{208}$Pb as well as its electric polarizability $\alpha_D$. Using the Bayesian analysis method, $E_{-1}$ and $\alpha_D$ are found to be mostly determined by the nuclear symmetry energy $E_{sym}$ at about $\rho^\star=0.05$ fm$^{-3}$ and the isovector nucleon effective mass $m_v^\star$ at the saturation density. At $90\%$ confidence level, we obtain $E_{sym}(\rho^\star) = 16.4 ^{+1.0} _{-0.9}$ MeV and $m_v^\star/m = 0.79 ^{+0.06} _{-0.06}$.
\end{abstract}

\maketitle

Understanding properties of nuclear interactions is one of the main goals of nuclear physics. So far the uncertainties mainly exist in isovector channels of nuclear interactions, and they manifest themselves in the isospin-dependent part of the nuclear matter equation of state (EOS) and the single-nucleon potential. The isospin-dependent part of the EOS, i.e., the symmetry energy $E_{sym}$, although still not well determined, is around 30 MeV at the saturation density $\rho_0$, while its density dependence characterized by the slope parameter $L=3\rho_0(dE_{sym}/d\rho)_{\rho_0}$ has recently been constrained within $L=58.7 \pm 28.1$ MeV~\cite{BAL13,Oer17} from various approaches. On the other hand, the momentum-dependence of the single-nucleon potential is generally characterized by the nucleon effective mass in the non-relativistic case, which is usually defined as that at the Fermi momentum in normal nuclear matter. Neutrons and protons may have different effective masses in isospin asymmetric nuclear matter, and the isospin splitting of the nucleon effective mass $m_{n-p}^\star \equiv (m_n^\star-m_p^\star)/m$ is even less constrained depending on the approaches used in the analysis (see, e.g., Table 2 in Ref.~\cite{Li18}). Both the symmetry energy and the isospin splitting of the nucleon effective mass have significant ramifications in nuclear reactions, nuclear structures, and nuclear astrophysics~\cite{Bar05,Ste05,Lat07,Li08,Li18}. They are related to each other through the Hugenholtz-Van Hove theorem~\cite{XuC10,BAL13}, and most isospin tracers are sensitive to both the $E_{sym}$ and the $m_{n-p}^\star$. This adds difficulties to accurately extracting the information of isovector nuclear interactions, unless a proper analysis using multiple observables is employed.

Observables of finite nuclei serve as good probes of nuclear interactions and nuclear matter properties at and below the saturation density. It has been found that useful information about isovector nuclear interactions can be extracted from the isovector giant dipole resonance (IVGDR)~\cite{Tri08,Rei10,Pie12,Vre12,Roc13b,Col14,Roc15,zhangzhen15,zhenghua16}, an oscillation mode in which neutrons and protons move collectively relative to each other in a nucleus. The centroid energy $E_{-1}$ of the IVGDR is a good probe of the $E_{sym}$ around and below the saturation density~\cite{Tri08,Col14}, while the product of the electric polarizability $\alpha_D$ and the $E_{sym}$ at the saturation density shows a good linear dependence on $L$~\cite{Roc13b,Roc15,Geb16}. Recently, it has been found that both the $E_{-1}$ and the $\alpha_D$ can be affected by $m_{n-p}^\star$ as well~\cite{Zha16,Kon17}, once the isoscalar nucleon effective mass is determined by the excitation energy of the isoscalar giant quadruple resonance (ISGQR)~\cite{Boh75,Boh79,Bla80,Klu09,Roc13a,Zha16,Kon17,Bon18}. Since both the $E_{-1}$ and the $\alpha_D$ are sensitive to the $E_{sym}$ and/or the $m_{n-p}^\star$, we employ the Bayesian analysis as in Ref.~\cite{Xie19} to extract $E_{sym}$ and $m_{n-p}^\star$ as well as their correlations, with properties of giant resonances calculated from the random-phase approximation (RPA) method based on the Skyrme-Hartree-Fock (SHF) model~\cite{Col13}.

The standard SHF functional originating from the following effective Skyrme interaction is used in the SHF-RPA calculation~\cite{Col13}
\begin{eqnarray}
v(\vec{r}_1,\vec{r}_2) &=& t_0(1+x_0P_\sigma)\delta(\vec{r}) \notag \\
&+& \frac{1}{2} t_1(1+x_1P_\sigma)[{\vec{k}'^2}\delta(\vec{r})+\delta(\vec{r})\vec{k}^2] \notag\\
&+&t_2(1+x_2P_\sigma)\vec{k}' \cdot \delta(\vec{r})\vec{k} \notag\\
&+&\frac{1}{6}t_3(1+x_3P_\sigma)\rho^\alpha(\vec{R})\delta(\vec{r}) \notag\\
&+& i W_0(\vec{\sigma}_1+\vec{\sigma_2})[\vec{k}' \times \delta(\vec{r})\vec{k}].
\end{eqnarray}
In the above, $\vec{r}=\vec{r}_1-\vec{r}_2$ and $\vec{R}=(\vec{r}_1+\vec{r}_2)/2$ are related to the positions of two nucleons $\vec{r}_1$ and $\vec{r}_2$, $\vec{k}=(\nabla_1-\nabla_2)/2i$ is the relative momentum operator and $\vec{k}'$ is its complex conjugate acting on the left, and $P_\sigma=(1+\vec{\sigma}_1 \cdot \vec{\sigma}_2)/2$ is the spin exchange operator. The parameters $t_0$, $t_1$, $t_2$, $t_3$, $x_0$, $x_1$, $x_2$, $x_3$, and $\alpha$ can be solved inversely from the macroscopic quantities~\cite{MSL0}, i.e., the saturation density $\rho_0$, the binding energy at the saturation density $E_0$, the incompressibility $K_0$, the isoscalar and isovector nucleon effective mass $m_s^\star$ and $m_v^\star$ at the Fermi momentum in normal nuclear matter, the symmetry energy and its slope parameter at the saturation density $E_{sym}^0$ and $L$, and the isoscalar and isovector density gradient coefficient $G_S$ and $G_V$. The analytical expressions of this inverse transformation are given by Eqs. (13)-(21) in Ref.~\cite{MSL0}, with no singularities from the empirical values of macroscopic quantities. The spin-orbit coupling constant is fixed at $W_0=133.3$ MeVfm$^5$. Although exploring the whole parameter space generally gives the best knowledge, it takes an incredible long time based on the available computational power. In the present practical study, we determine the value of $m_s^\star$ from the excitation energy of the ISGQR in $^{208}$Pb, and calculate the posterior probability distribution functions (PDFs) of the isovector interaction parameters, i.e., $E_{sym}^0$, $L$, and $m_v^\star$, through the Bayesian analysis, while keeping the values of the other macroscopic quantities the same as the empirical ones from the MSL0 interaction~\cite{MSL0}, since the $E_{-1}$ and $\alpha_D$ have been shown to be most sensitive to $E_{sym}^0$, $L$, and $m_v^\star$~\cite{Tri08,Col14,Roc13b,Roc15,Geb16,Zha16,Kon17,Zha14}.

The operators for the IVGDR and ISGQR are chosen respectively as
\begin{equation}
\hat{F}_{\rm 1M} = \frac{N}{A} \sum_{i=1}^Z r_i Y_{\rm 1M}(\hat{r}_i) - \frac{Z}{A} \sum_{i=1}^N r_i Y_{\rm 1M}(\hat{r}_i), \label{QIVGDR}
\end{equation}
and
\begin{equation}
\hat{F}_{\rm 2M} = \sum_{i=1}^A r_i^2 Y_{\rm 2M}(\hat{r}_i),
\end{equation}
where $N$, $Z$, and $A$ are respectively the neutron, proton, and nucleon numbers in a nucleus, $r_i$ is the coordinate of the $i$th nucleon with respect to the center-of-mass of the nucleus, and $Y_{\rm 1M}(\hat{r}_i)$ and $Y_{\rm 2M}(\hat{r}_i)$ are the spherical Bessel functions, with the magnetic quantum number $M$ degenerate in spherical nuclei. Using the RPA method~\cite{Col13}, the strength function
\begin{equation}
S(E) = \sum_\nu |\langle \nu|| \hat{F}_{J}  || \tilde{0} \rangle |^2 \delta(E-E_\nu)
\end{equation}
of a nucleus resonance can be obtained, where the square of the reduced matrix element $|\langle \nu|| \hat{F}_{J}  || \tilde{0} \rangle |$ represents the transition probability from the ground state $| \tilde{0} \rangle $ to the excited state $| \nu \rangle$. The moments of the strength function can then be calculated from
\begin{equation}
m_k = \int_0^\infty dE E^k S(E).
\end{equation}
The centroid energy $E_{-1}$ of the IVGDR and the electric polarizability $\alpha_D$ can be obtained from the moments of the strength function through the relation
\begin{eqnarray}
E_{-1} &=& \sqrt{m_1/m_{-1}}, \\
\alpha_D &=& \frac{8\pi e^2}{9} m_{-1}.
\end{eqnarray}
The moments are not used in the ISGQR analysis, since the excitation energy $E_x$ is the peak energy of the strength function to be compared with the corresponding experimental result.

The Bayesian analysis is used to obtain the PDFs of model parameters from the experimental data. Such PDFs can be formally calculated from the Bayes' theorem
\begin{equation}
P(M|D) = \frac{P(D|M)P(M)}{\int P(D|M)P(M)dM}.
\end{equation}
In the above, $P(M|D)$ is the posterior probability for the model $M$ given the data set $D$, $P(D|M)$ is the likelihood function or the conditional probability for a given theoretical model $M$ to predict correctly the data $D$, and $P(M)$ denotes the prior probability of the model $M$ before being confronted with the data. The denominator of the right-hand side of the above equation is the normalization constant. For the prior PDFs, we choose the model parameters $p_1=E_{sym}^0$ uniformly within $25 \sim 35$ MeV, $p_2=L$ uniformly within $0 \sim 120$ MeV, and $p_3=m_v^\star/m$ uniformly within $0.5 \sim 1$, with $m$ being the bare nucleon mass. Comparing with the experimental data, the binding energies and charge radii of $^{208}$Pb and other nuclei are deviated by only a few percent at most, by changing the values of $E_{sym}^0$, $L$, and $m_v^\star/m$ within their prior ranges~\cite{MSL0}, showing that the Bayesian analysis is exploring the reasonable parameter space. The theoretical results of $d^{th}_1=E_{-1}$ and $d^{th}_2=\alpha_D$ from the SHF-RPA method are used to calculate the likelihood of these model parameters with respect to the corresponding experimental data $d^{exp}_1$ and $d^{exp}_2$ according to
\begin{eqnarray}
&&P[D(d_{1,2})|M(p_{1,2,3})] \notag\\
&=& \frac{1}{2\pi \sigma_1 \sigma_2} \exp\left[-\frac{(d^{th}_1-d^{exp}_1)^2}{2\sigma_1^2}-\frac{(d^{th}_2-d^{exp}_2)^2}{2\sigma_2^2}\right], \label{llh}
\end{eqnarray}
where $\sigma_{1,2}$ denote the widths of the likelihood function. The posterior PDF of a single model parameter $p_i$ is given by
\begin{equation}
P(p_i|D) = \frac{\int P(D|M) P(M) \Pi_{j\ne i} dp_j}{\int P(D|M) P(M) \Pi_{j} dp_j},
\end{equation}
while the correlated PDF of two model parameters $p_i$ and $p_j$ is given by
\begin{equation}
P[(p_i,p_j)|D] = \frac{\int P(D|M) P(M) \Pi_{k\ne i,j} dp_{k}}{\int P(D|M) P(M) \Pi_{k} dp_k}.
\end{equation}
The calculation of the posterior PDFs is based on the Markov-Chain Monte Carlo (MCMC) approach using the Metropolis-Hastings algorithm~\cite{Met53,Has70}. Since the MCMC process does not start from an equilibrium distribution, initial samples in the so-called burn-in period have to be thrown away.

The mean values of the experimentally measured excitation energy $E_x=10.9$ MeV~\cite{Roc13a} of the ISGQR in $^{208}$Pb can be reproduced by using $m_s^\star/m=0.83$ approximately independent of other macroscopic quantities, whose values remain unchanged as those from the MSL0 interaction~\cite{MSL0} for the IVGDR analysis. The small experimental error bars of $E_x$ are neglected in the present study. For the given $m_s^\star/m$, the experimental results of the centroid energy $E_{-1}=13.46$ MeV of the IVGDR from photoneutron scatterings~\cite{IVGDRe}, and the electric polarizability $\alpha_D=19.6 \pm 0.6$ fm$^3$ from polarized proton inelastic scatterings~\cite{Tam11} and with the quasi-deuteron excitation contribution subtracted~\cite{Roc15}, are used in the Bayesian analysis. Larger $\sigma_1$ and $\sigma_2$ values are used to evaluate the likelihood function in the early stage of the MCMC process, in order to accelerate the convergence procedure, and it gradually decreases to the $1\sigma$ error from the experimental measurement, after which the results are analyzed. An artificial $1\sigma$ error of $0.1$ MeV for the well-determined $E_{-1}$ value of the IVGDR is used in the analysis after convergence.

\begin{figure}[ht]
	\includegraphics[scale=0.25]{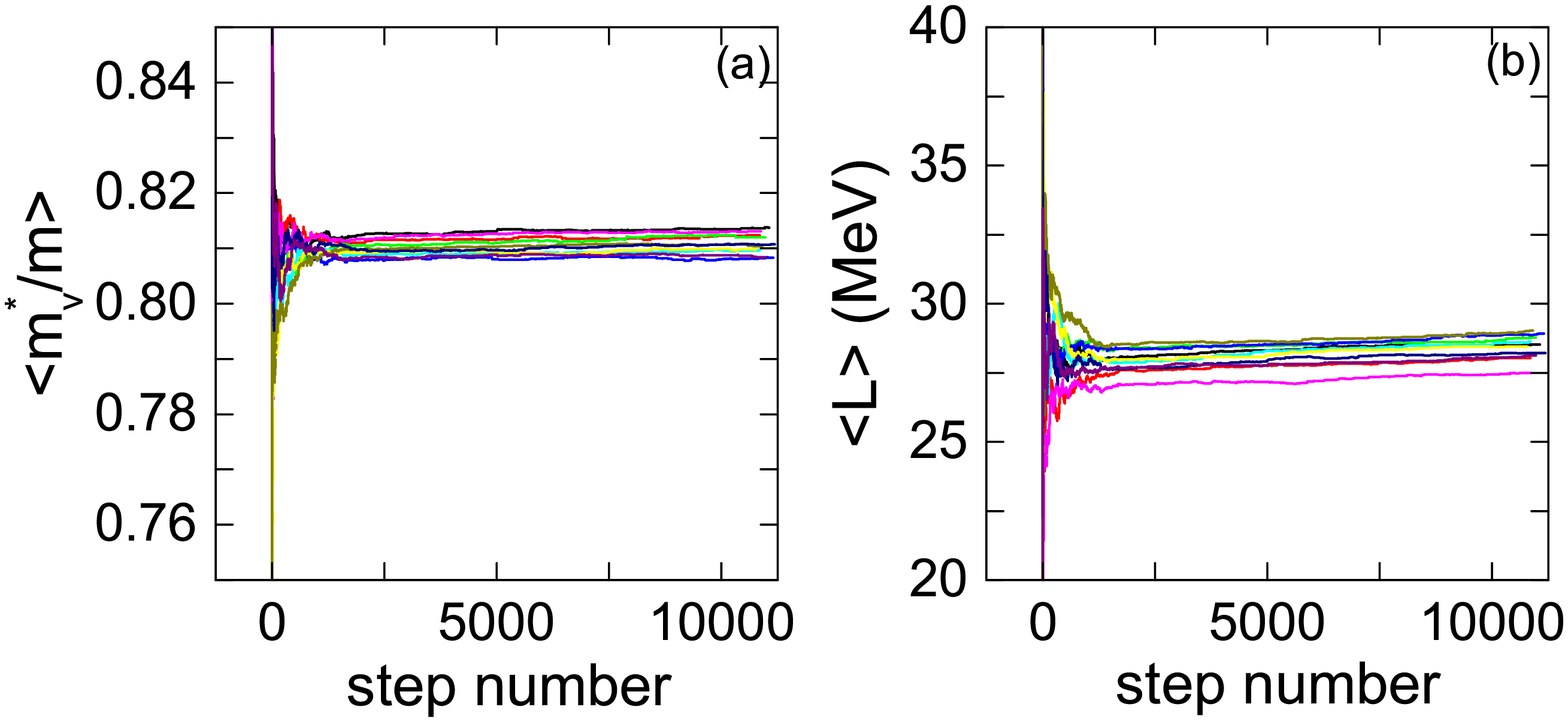}\\
	\includegraphics[scale=0.3]{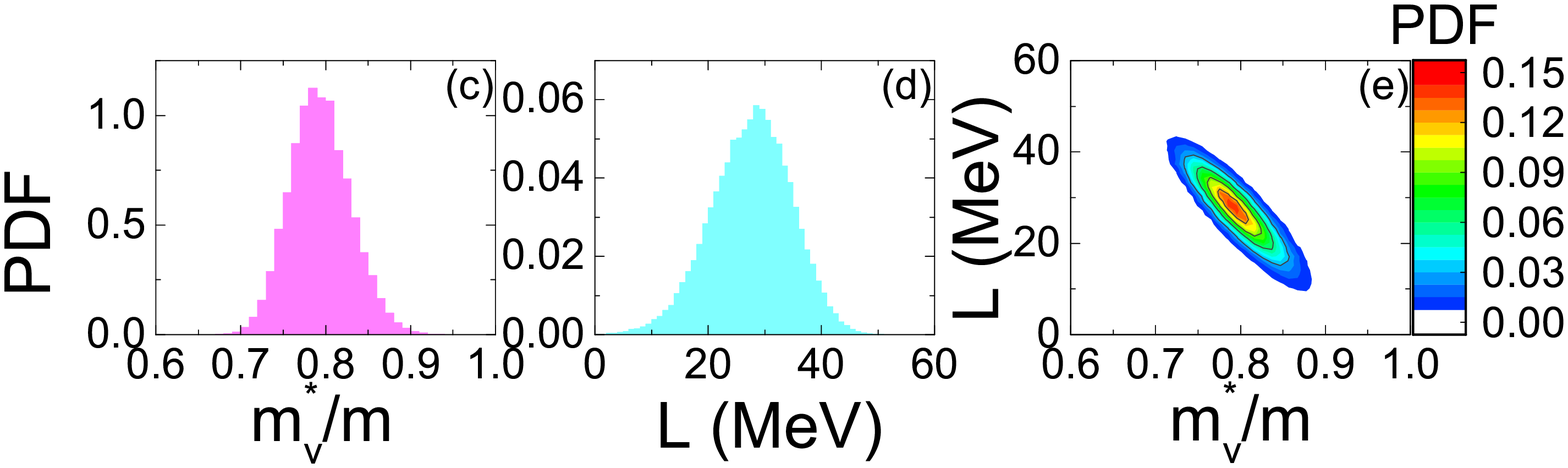}
	\caption{(Color online) Upper: Mean values of $m_v^\star/m$ (a) and $L$ (b) as a function of the step number at a fixed $E_{sym}^0=30$ MeV for 10 parallel runs; lower: The PDFs of $m_v^\star/m$ (c) and $L$ (d) as well as their correlations (e) at a fixed $E_{sym}^0=30$ MeV. } \label{fig1}
\end{figure}

By fixing $E_{sym}^0=30$ MeV, we first study the posterior PDFs of $m_v^\star/m$ and $L$, and their mean values as a function of the step number are plotted in the upper panels of Fig.~\ref{fig1}, for 10 parallel runs. It is seen that the convergence is generally reached after a few thousand steps. The PDFs are thus from analyzing the results after about 2000 steps, and until 10000 steps there are totally about 800 accepted data samples for each run. The posterior PDFs of $m_v^\star/m$ and $L$ as well as their correlations are plotted in the lower panels of Fig.~\ref{fig1}. It is seen that the PDF of $m_v^\star/m$ peaks around 0.8, while that of $L$ peaks around 30 MeV. The anticorrelation between $m_v^\star/m$ and $L$ for a fixed $E_{sym}^0$ is observed. A narrower anticorrelation is expected to be observed by using a smaller artificial $1\sigma$ error for $E_{-1}$, but with the slope of the anticorrelation unchanged. Although the $L$ values from the above Bayesian analysis at a fixed $E_{sym}^0=30$ MeV are small compared with the average ones extracted from various approaches in Refs.~\cite{BAL13,Oer17}, they are consistent with the IVGDR result in Ref.~\cite{Tri08} (see also the "GDR" band in Fig.~1 of Ref.~\cite{Lat14}).

\begin{figure}[ht]
	\includegraphics[scale=0.3]{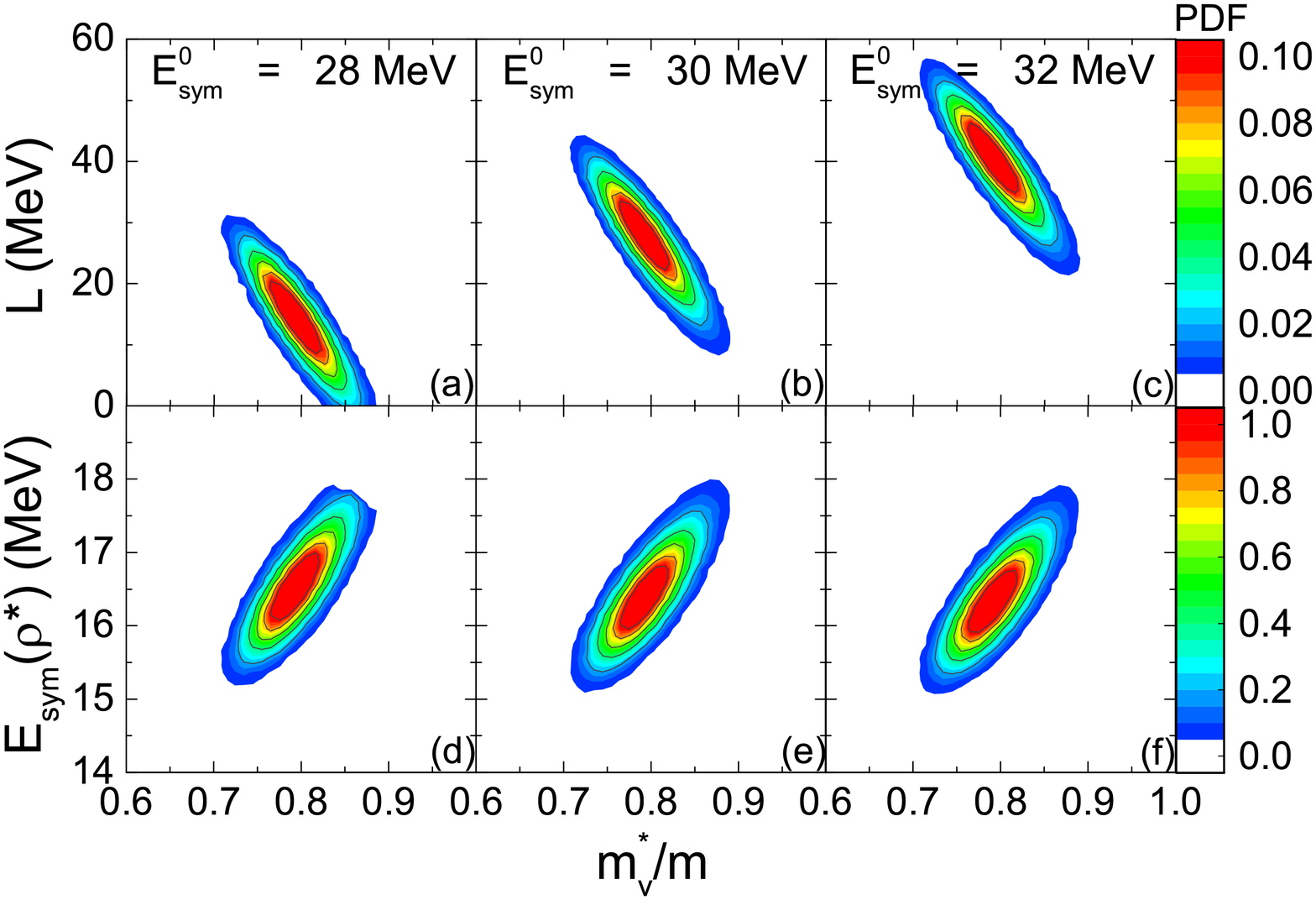}
	\caption{(Color online) PDFs in the $m_v^\star/m-L$ plane (upper panels) and in the $m_v^\star/m-E_{sym}(\rho^\star)$ plane (lower panels) at $E_{sym}^0=28$, 30, and 32 MeV, with $\rho^\star = 0.05$ fm$^{-3}$. } \label{fig2}
\end{figure}

\begin{figure}[ht]
	\includegraphics[scale=0.3]{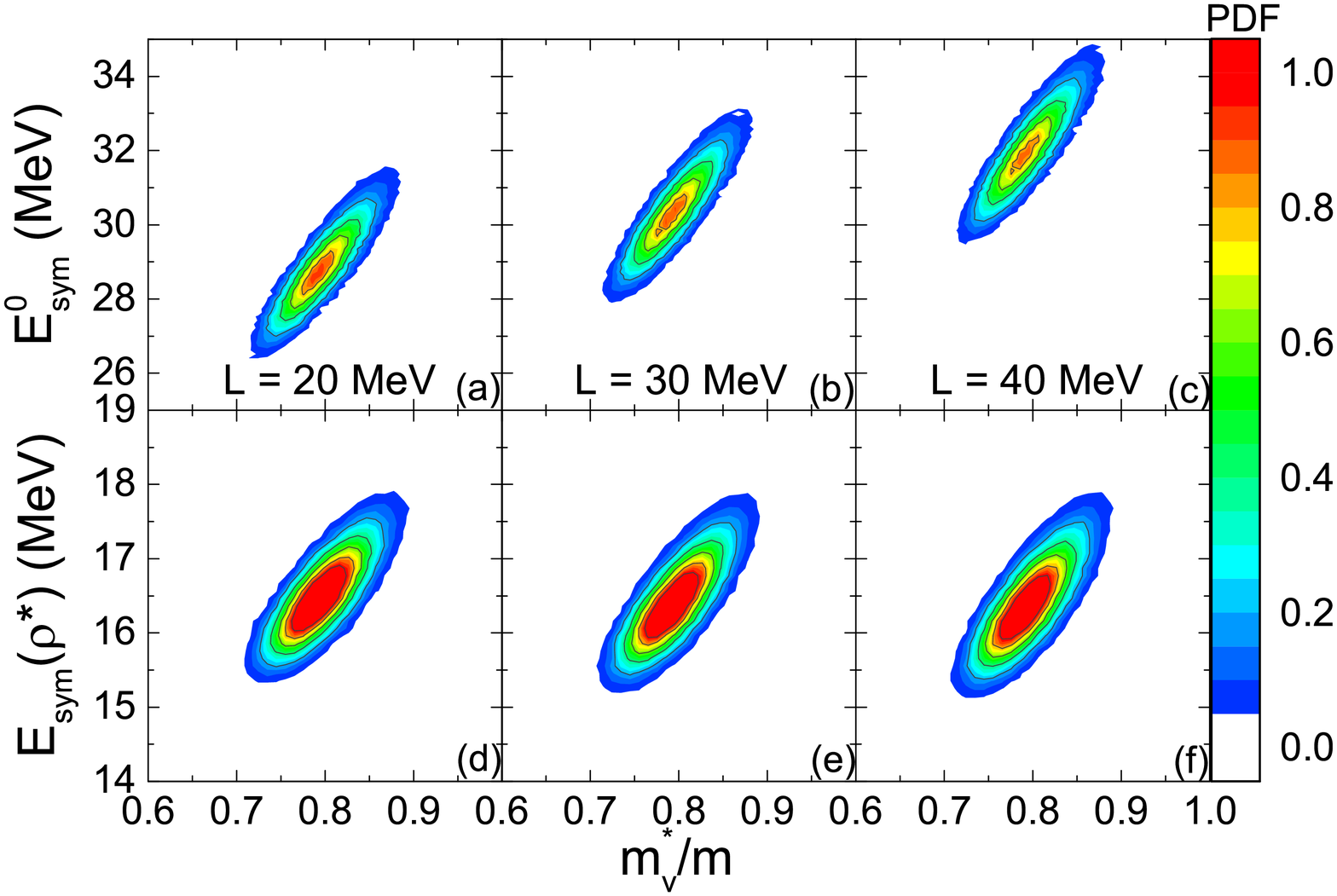}
	\caption{(Color online) PDFs in the $m_v^*-E_{sym}^0$ plane (upper panels) and in the $m_v^\star/m-E_{sym}(\rho^\star)$ plane (lower panels) at $L=20$, 30, and 40 MeV, with $\rho^\star = 0.05$ fm$^{-3}$. } \label{fig3}
\end{figure}

The uncertainties of $E_{sym}^0$ are expected to affect the extracted PDFs of $m_v^\star/m$ and $L$ as well as their correlations. The upper panels of Fig.~\ref{fig2} compare the correlations between $m_v^\star/m$ and $L$ at $E_{sym}^0=28$, 30, and 32 MeV, respectively. For a larger $E_{sym}^0$, the PDF moves to the upper side of the figure with a larger $L$ value, while the PDF of $m_v^\star/m$ as well as the anticorrelation between $m_v^\star/m$ and $L$ remain almost unchanged. Similarly, it is also possible to study the correlation between $m_v^\star/m$ and $E_{sym}^0$ at a fixed $L$, and the results are shown in the upper panels of Fig.~\ref{fig3}. The positive correlation between $m_v^\star/m$ and $E_{sym}^0$ for a given $L$ is observed. For a larger $L$, the PDF moves to the upper side of the figure with a larger $E_{sym}^0$ value, while the shape of PDF remains almost the same.

\begin{figure}[ht]
	\includegraphics[scale=0.3]{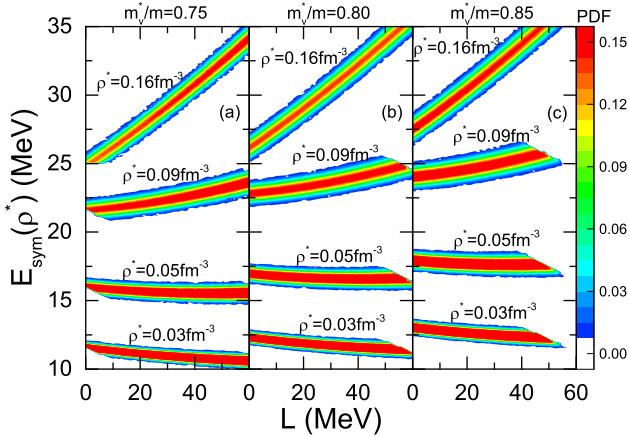}
	\caption{(Color online) PDFs in the $L-E_{sym}(\rho^\star)$ plane at $m_v^\star/m=0.75$ (a), 0.80 (b), and 0.85 (c) for different values of $\rho^\star$.} \label{fig4}
\end{figure}

Inspired by the regular behaviors observed above, we have further studied the correlation between $L$ and $E_{sym}(\rho^\star)$ for different $\rho^\star$ at a fixed $m_v^\star/m$ in Fig.~\ref{fig4}. For $\rho^\star=0.16$ fm$^{-3}$, a nearly linear and positive correlation between $L$ and $E_{sym}^0=E_{sym}(\rho^\star)$ is observed. Similar linear relations were extracted in Ref.~\cite{Roc15} from the neutron-skin thickness and $\alpha_D$ for various nuclei. This positively linear correlation between $L$ and $E_{sym}^0$ is consistent with the behaviors observed in the upper panels of Figs.~\ref{fig2} and \ref{fig3}, showing that properties of the IVGDR are sensitive to the $E_{sym}(\rho^\star)$ at $\rho^\star$ other than the saturation density. The latter can be calculated from the SHF functional, with given values of $L$, $m_v^\star/m$, $E_{sym}^0$, and other default quantities from the MSL0 interaction. From positive to slightly negative correlations between $L$ and $E_{sym}(\rho^\star)$ are observed with the decreasing value of $\rho^\star$. It is interesting to see that for $\rho^\star=0.05$ fm$^{-3}$ values of $E_{sym}(\rho^\star)$ become approximately uncorrelated with $L$, showing that $E_{-1}$ and $\alpha_D$ are most sensitive to the symmetry energy at that density. This is consistent with the conclusion from Ref.~\cite{Zha14} that $\alpha_D$ of $^{208}$Pb is strongly correlated with the symmetry energy at about $\rho_0/3$. The observed cutoffs in the PDFs are due to the choice of the prior distribution, i.e., $E_{sym}^0$ within $25 \sim 35$ MeV. The correlations are shifted at different fixed $m_v^\star/m$ values, while the strong correlation between properties of the IVGDR in $^{208}$Pb and $E_{sym}$ at $\rho^\star=0.05$ fm$^{-3}$ remains robust.

The above finding shows that the sensitivity of IVGDR properties to the $E_{sym}$ at $\rho^\star=0.05$ fm$^{-3}$ instead of $L$ or $E_{sym}^0$ as well as $m_v^\star/m$ is a robust feature based on the Bayesian analysis. To further confirm this finding, we have calculated $E_{sym}(\rho^\star)$ from $L$, $m_v^\star/m$, $E_{sym}^0$, and other default quantities from the MSL0 interaction, and replotted the upper panels of Fig.~\ref{fig2} and Fig.~\ref{fig3}. For different values of $L$ and $E_{sym}^0$, the resulting correlations in the $m_v^\star/m-E_{sym}(\rho^\star)$ plane are displayed in the lower panels of the corresponding figures. It is seen that these PDFs are almost the same and thus approximately independent of $L$ and $E_{sym}^0$.

\begin{figure}[hbt]
	\includegraphics[scale=0.4]{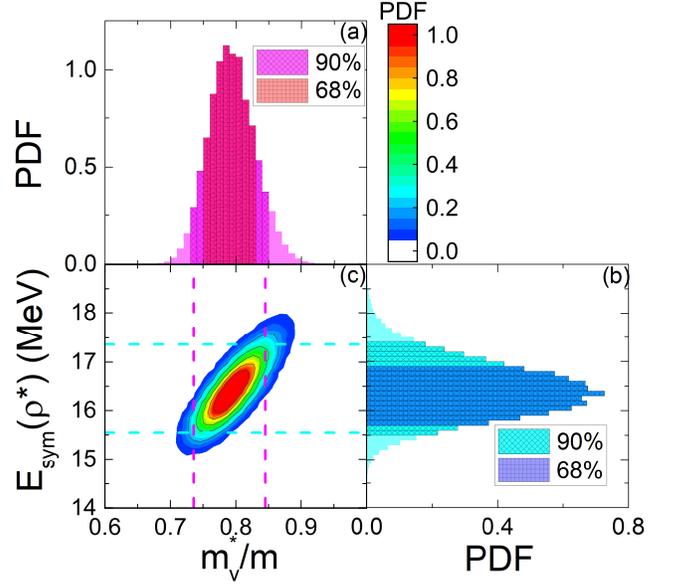}
	\caption{(Color online) The PDFs of $m_v^\star/m$ (a) and $E_{sym}$ at $\rho^\star = 0.05$ fm$^{-3}$ (b) as well as their correlations (c) roughly independent of $L$ and $E_{sym}^0$. } \label{fig5}
\end{figure}

The final resulting PDFs of $m_v^\star/m$ and $E_{sym}(\rho^\star)$ as well as their correlations in the present study are shown in Fig.~\ref{fig5}. It is seen that $m_v^\star/m$ are positively correlated with $E_{sym}(\rho^\star)$ under the constraints of $E_{-1}$ and $\alpha_D$. We obtain $m_v^\star/m = 0.79 ^{+0.04} _{-0.04}$ and $E_{sym}(\rho^\star) = 16.4 ^{+0.5} _{-0.7}$ MeV at $68\%$ confidence level, and $m_v^\star/m = 0.79 ^{+0.06} _{-0.06}$ and $E_{sym}(\rho^\star) = 16.4 ^{+1.0} _{-0.9}$ MeV at $90\%$ confidence level. The $90\%$ confidence interval of $m_v^\star/m$ together with $m_s^\star/m=0.83$ leads to the neutron-proton effective mass splitting $m_{n-p}^\star \approx 0.084 ^{+0.143} _{-0.123} \delta$ in normal nuclear matter with the isospin asymmetry $\delta$.

In conclusion, we have studied the IVGDR in $^{208}$Pb from the random-phase approximation method based on the Skyrme-Hartree-Fock model, and employed the Bayesian analysis to extract the posterior PDFs of isovector parameters from the centroid energy of the IVGDR and the electric polarizability. Inspired by the similar shape of the PDFs for a given symmetry energy at the saturation density $E_{sym}^0$ or the slope parameter $L$ of the symmetry energy as well as the linear correlation between $L$ and $E_{sym}^0$, we found that properties of IVGDR are mostly determined by the positive correlation between the symmetry energy at $\rho^\star \approx 0.05$ fm$^{-3}$ and the isovector nucleon effective mass $m_v^\star$, but not directly by $L$ and $E_{sym}^0$. Moreover, $m_v^\star/m = 0.79 ^{+0.06} _{-0.06}$ at the saturation density and $E_{sym}(\rho^\star) = 16.4 ^{+1.0} _{-0.9}$ MeV are obtained at $90\%$ confidence level from the present study.

JX acknowledges the National Natural Science Foundation of China under Grant No. 11922514. ZZ acknowledges the National Natural Science Foundation of China under Grant No. 11905302. WJX acknowledges the National Natural Science Foundation of China under Grant No. 11505150. BAL acknowledges the U.S. Department of Energy, Office of Science, under Award Number DE-SC0013702, the CUSTIPEN (China-U.S. Theory Institute for Physics with Exotic Nuclei) under the US Department of Energy Grant No. DE-SC0009971.



\begin{thebibliography}{99}



\bibitem{BAL13} B. A. Li and X. Han, Phys. Lett. B \textbf{727}, 276 (2013).

\bibitem{Oer17} M. Oertel, M. Hempel, T. Kl\"ahn, and S. Typel, Rev. Mod. Phys. \textbf{89}, 015007 (2017).

\bibitem{Li18} B. A. Li, B. J. Cai, L. W. Chen, and J. Xu, Prog. Part. Nucl. Phys. \textbf{99}, 29 (2018).

\bibitem{Bar05} V. Baran, M. Colonna, V. Greco, and M. Di Toro, Phys. Rep \textbf{410}, 335 (2005).

\bibitem{Ste05} A. W. Steiner, M. Prakash, J. M. Lattimer, and P. J. Ellis, Phys. Rep. \textbf{411}, 325 (2005).

\bibitem{Lat07} J. M. Lattimer and M. Prakash, Phys. Rep. \textbf{442}, 109 (2007).

\bibitem{Li08} B. A. Li, L. W. Chen, and C. M. Ko, Phys. Rep. \textbf{464}, 113 (2008).

\bibitem{XuC10} C. Xu, B. A. Li, and L. W. Chen, Phys. Rev. C \textbf{82}, 054607 (2010).

\bibitem{Tri08} L. Trippa, G. Col\`o, and E. Vigezzi, Phys. Rev. C \textbf{77}, 061304(R) (2008).

\bibitem{Rei10} P.-G. Reinhard and W. Nazarewicz, Phys. Rev. C \textbf{81}, 051303(R) (2010).

\bibitem{Pie12} J. Piekarewicz, B. K. Agrawal, G. Col\`o, W. Nazarewicz, N. Paar, P.-G. Reinhard, X. Roca-Maza, and D. Vretenar,
     Phys. Rev. C \textbf{85}, 041302(R) (2012).

\bibitem{Vre12} D. Vretenar, Y. F. Niu, N. Paar, and J. Meng, Phys. Rev. C \textbf{85}, 044317 (2012).

\bibitem{Roc13b} X. Roca-Maza, M. Brenna, G. Col\`o, M. Centelles, X. Vi\~nas, B. K. Agrawal, N. Paar, D. Vretenar, and J. Piekarewicz, Phys. Rev. C \textbf{88}, 024316 (2013).

\bibitem{Col14} G. Col\`o, U. Garg, and H. Sagawa, Eur. Phys. J. A \textbf{50}, 26 (2014).

\bibitem{Roc15} X. Roca-Maza, X. Vi\~nas, M. Centelles, B. K. Agrawal, G. Col\`o, N. Paar, J. Piekarewicz, and D. Vretenar, Phys. Rev. C \textbf{92}, 064304 (2015).

\bibitem{zhangzhen15} Z. Zhang and L. W. Chen, Phys. Rev. C \textbf{93}, 031301(R) (2015).

\bibitem{zhenghua16} H. Zheng, S. Burrello, M. Colonna, and V. Baran, Phys. Rev. C \textbf{94}, 014313 (2016). 

\bibitem{Geb16} E. Gebrerufael, A. Calci, and R. Roth, Phys. Rev. C \textbf{93}, 031301(R) (2016).

\bibitem{Zha16} Z. Zhang and L. W. Chen, Phys. Rev. C \textbf{93}, 034335 (2016).

\bibitem{Kon17} H. Y. Kong, J. Xu, L. W. Chen, B. A. Li, and Y. G. Ma, Phys. Rev. C \textbf{95}, 034324 (2017).

\bibitem{Boh75} A. Bohr and B. R. Mottelson, {\it Nuclear Stucture}, Vols. I and II (W. A. Benjamin Inc., Reading, MA, 1975).

\bibitem{Boh79} O. Bohigas, A. M. Lane, and J. Martorell, Phys. Rep. \textbf{51}, 267 (1979).

\bibitem{Bla80} J.-P. Blaizot, Phys. Rep. \textbf{64}, 171 (1980).

\bibitem{Klu09} P. Kl\"upfel, P.-G. Reinhard, T. J. B\"urvenich, and J. A. Maruhn, Phys. Rev. C \textbf{79}, 034310 (2009).

\bibitem{Roc13a} X. Roca-Maza, M. Brenna, B. K. Agrawal, P. F. Bortignon, G. Col\`o, L. G. Cao, N. Paar, and D. Vretenar, Phys. Rev. C \textbf{87}, 034301 (2013).

\bibitem{Bon18} G. Bonasera, M. R. Anders, and S. Shlomo, Phys. Rev. C \textbf{98}, 054316 (2018).

\bibitem{Xie19} W. J. Xie and B. A. Li, Astro. Phys. J. \textbf{883}, 174 (2019).

\bibitem{Col13} G. Col\`o, L. Cao, N. Van Gia, and L. Capelli, Com. Phys. Com. \textbf{184}, 142 (2013).

\bibitem{MSL0} L. W. Chen, B. A. Li, C. M. Ko, and J. Xu, Phys. Rev. C \textbf{82}, 024321 (2010).

\bibitem{Zha14} Z. Zhang and L. W. Chen, Phys. Rev. C \textbf{90}, 064317 (2014).

\bibitem{Met53} N. Metropolis, A. W. Rosenbluth, M. N. Rosenbluth, and A. H. Teller, J. Chem. Phys. \textbf{21}, 1087 (1953).

\bibitem{Has70} W. K. Hastings, Biometrika \textbf{57}, 97 (1970).




\bibitem{IVGDRe} S. S. Dietrich and B. L. Berman, At. Data Nucl. Data Tables \textbf{38}, 199 (1988).

\bibitem{Tam11} A. Tamii, I. Poltoratska, P. vonNeumann-Cosel, {\it et al.}, 
    Phys. Rev. Lett. \textbf{107}, 062502 (2011).







\bibitem{Lat14} J. M. Lattimer and A. W. Steiner, Eur. Phys. J. A \textbf{50}, 40 (2014).
\end{thebibliography}
\end{document}